\documentclass[pra,aps,preprint]{revtex4}
\usepackage{amssymb}

\usepackage{graphicx}

\begin{document}

\title{Localization of Macroscopic Object Induced by the Factorization of Internal
Adiabatic Motion}
\author{\textsc{C.P.Sun$^{1}$,X.F.Liu$^{1,2}$ , D.L.Zhou$^{1}$ and S.X.Yu$^{1}$ }}

\address{$^{(1)} $ Institute of Theoretical Physics, Chinese Academy of Sciences,
Beijing 100080, China\\$^{(2)} $ Department of Mathematics, Peking
University,Beijing, 100871,China} \

\begin{abstract}
To account for the phenomenon of quantum decoherence of a macroscopic
object, such as the localization and disappearance of interference, we
invoke the adiabatic quantum entanglement between its collective states(such
as that of the center-of-mass (C.M)) and its inner states based on our
recent investigation. Under the adiabatic limit that motion of C.M dose not
excite the transition of inner states, it is shown that the wave function of
the macroscopic object can be written as an entangled state with correlation
between adiabatic inner states and quasi-classical motion configurations of
the C.M. Since the adiabatic inner states are factorized with respect to
each parts composing the macroscopic object, this adiabatic separation can
induce the quantum decoherence. This observation thus provides us with a
possible solution to the Schroedinger cat paradox. \vspace{1cm}

\end{abstract}

\maketitle

\section{Introduction}

It is common sense that a macroscopic object should be localized in certain
spatial domain. However, problem will appear if one directly use quantum
mechanics to describe the motion of a free macroscopic object with spatial
localization .This issue originated from the correspondence between Einstein
and Born [1]. They observed that, in a spatially-localized state, generally
a macroscopic object can only be described by a time-dependent localized
wave packet, which is a coherent superposition of the eigen-states of the
center-of-mass Hamiltonian $H_0=p^2/2M$. If the macroscopic object is
regarded as a heavy particle of a large mass $M$ , its initial state $%
|\varphi \rangle $ should be a very narrow wave packet of width $a$. Since
the wave packet spreads in evolution by the law
\begin{equation}
w(t)=a\sqrt{1+\frac{t^2}{4M^2a^4},}
\end{equation}
where $w\left( t\right) $ stands for the width of the wave packet, the
spreading of an initially well localized wave packet can be reasonably
ignored for very large mass. This seems to give a solution to the
localization problem of the macroscopic object. But Einstein argued that the
superposition of two narrow wave packets is no longer narrow with respect to
the macro-coordinate, and on the other hand, it is still a possible state of
the macroscopic object. So a contradiction to the superposition principle
arises because of the requirement that the wave packet of a macroscopic
object should be narrow [1].

To solve this problem, Wigner [2], Joos and Zeh [3] propose the so called
scattering -induced -decoherence mechanism (or WJZ mechanism): scattering of
photons and atoms off a macroscopic object records the information of its
position to form a quantum measurement about the position. Then the
interference terms between different paths of the macroscopic object are
destroyed by the generalized ''which-way''detection in association with
scattering. In fact , in quantum measurement process, wave packet collapse
(WPC, also called von Neumann's projection ) physically resembles the
disappearance of interference pattern in Young's two-slit experiment in the
presence of a ``which-way'' detector. Associated with the wave-particle
duality, this phenomenon of losing quantum coherence is referred to as
quantum decoherence [4]: before a measurement to observe ``which-way'' the
particle actually takes, the quantum particle seems to move from one point
to another along several different ways simultaneously. This just reflects
the wave feature of a quantum particle. The detection of ``which-way'' means
a probe for the particle feature, which leads to the disappearance of wave
feature or quantum decoherence.I indeed, based on the Bragg's reflection of
cold atoms and the electronic Aharonov-Bohm interference with a quantum
point contact , the most recent experiments [5,6] shows that Schroedinger's
concept of entangled state, rather than the unavoidable measurement
distribution, is crucial for the wave-particle duality in this ``which-way''
detection. It is also pointed out that, similar gedenken experiments using
photon and neutron have been considered before [7,8].

With these experiments and theory, it seems to be concluded that there does
not exist the coherent superposition of states of a macroscopic object due
to the quantum decoherence resulting from its coupling to an external
environment as a generalized detector. But a natural question arises : If a
macroscopic object, such as the famous Schroedinger cat, is completely
isolated from any external environment, can its quantum coherence be
maintained to realize macroscopic superposition states? To answer this
question, one must consider the influences of the inner particles as the
so-called ''internal environment''[9] constituting the macroscopic object.
Most recently, as a new way round the quantum coherence of macroscopic
object , a novel experiment was presented to observe the matter wave
interference of C$_{60}$ molecules by diffraction at a material absorption
grating [10]. For the purpose of observing quantum coherence this molecule
is more massive than anything else that has been studied in this way before.
As a large molecule at a high temperature, C$_{60}$ contains atoms in
continual motion which remains coherent while the molecule is passing
through the slits. However, based on the above mentioned experiment, it
might be possible to set up decoherence experiments so long as one can find
a new way to effectively record the ''which-way '' information of C$_{60}$,
which is manifested by the its radiated infrared photons. Actually, there
still appears the coherent superposition of the macroscopic states in
certain extreme cases such as in superconductivity and Bose-Einstein
condensation [11], but these are not in our case since these
macroscopically-quantum phenomenon must require each part of the macroscopic
object has a same phase in evolution.

Motivated by these achievements both in theoretical and experimental
aspects, we will show that the conception of adiabatic quantum entanglement(
most recently proposed in refs.[12] based on the Born-Oppenhemeir (BO)
approximation ) is mainly responsible for the decoherence phenomenon of the
macroscopic object, such as the localization and disappearance of
interference. This kind of quantum entanglement occurs between the states of
the center-of-mass (C.M) of the macroscopic object and its inner states . In
fact, when the motion of C.M dose not excite the transition of inner states,
the wave function of the macroscopic object can be adiabatically factorized
with correlation between adiabatic inner states and quasi-classical motion
configurations of the C.M. By this correlation or entanglement, the spatial
localization of a macroscopic object can be explained and the dilemma of the
Schroedinger cat can be resolved in a natural way.

\section{Adiabatic Entanglement and WJZ mechanism}

In this section, based on the idea of the adiabatic entanglement, we
incorporate the WJZ mechanism and the relevant studies in the dynamic
theories of quantum measurement [13-16]developed by many people, including
one (CPS) of the authors .

In the WJZ mechanism, the ''which-way'' information of the macroscopic
object is recorded through the quantum entanglement formed by the scattering
of atoms or photos forming so-called environment.. Let $x$ and $q$ be,
respectively, the collective position (C.M) of a macroscopic object and
environment variables. The collective Hamiltonian is $H_s=p^2/2M$ and the
canonical commutation relations are $[x,p]=i,[x,q]=0.$ To study how
different positions of the macroscopic object entangle with environment
(scattering atoms,photons, et.al.) , we suppose that the total system is
initially in a product state
\begin{equation}
|\Psi _x(t=0)\rangle =|x\rangle \otimes |\phi \rangle
\end{equation}
where the first component $|x\rangle $ is the eigen-state of the collective
position operator $x$ while $|\phi \rangle $ is an arbitrary pure state of
the environment. Usually, the collective motion acts on the environment in
certain ways and the back-action of the environment can not be neglected
physically for a large macroscopic object . So this generic interaction can
not produce an ideal entanglement between the collective position of the
macroscopic object and the states of environment. By an argument by Joos and
Zeh [4] , one observes that only when the back-action is negligibly small,
can the interaction between the collective and environment states realize a
''measurement-like process'':
\begin{equation}
|x\rangle \otimes |\phi \rangle \rightarrow U(t)|x\rangle \otimes |\phi
\rangle =|x(t)\rangle \otimes S(x;t)|\phi \rangle
\end{equation}
Here, $U(t)$ is the total evolution matrix, $|x(t)\rangle $ $%
=U_0(t)|x\rangle $ represents the free evolution in the absence of the
coupling to the environment; and $S(x,t)$, acting on the environment states,
denotes the effective $S-matrix$ parameterized by the collective position $x$
of the macroscopic object. If the collective motion is initially described
by a wave packet $|\varphi \rangle =\int \varphi (x)|x\rangle dx,$then the
continuous quantum entangling state
\[
|\Phi \rangle =\int \varphi (x)|x(t)\rangle \otimes S(x;t)|\phi \rangle dx
\]
defines the reduced density matrix
\begin{equation}
\rho (x,x^{\prime },t)=\varphi (x,t)\varphi ^{*}(x^{\prime },t)\langle \phi
|S^{\dagger }(x^{\prime };t)S(x;t)|\phi \rangle
\end{equation}
of the macroscopic object. Considering the translational invariance of the
scattering process, Joos and Zeh showed that, , the off-diagonal terms take
the following form
\begin{equation}
F(x,x^{\prime })=\langle \phi |S^{\dagger }(x^{\prime };t)S(x;t)|\phi
\rangle \sim \exp (-\Lambda t|x-x^{\prime }|^2).
\end{equation}
in $x-representation.$This means the decoherence factor is a damping
function with the localization rate $\Lambda $ , which depends on the total
cross section.

In general, a quantum entangled state for a macroscopic object coupling to
an external environment reads $|\Psi \rangle =\sum_nC_n|M_n\rangle \otimes
|E_n\rangle .$ It involves a correlation between the states $|M_n\rangle $
of the macroscopic object and the states $|E_n\rangle $ of the environment.
The interference pattern
\begin{eqnarray}
p(x) &=&\sum_m|\langle E_m,x|\Psi \rangle |^2=\sum_n|C_n|^2|M_n(x)|^2
\nonumber \\
&&+\sum_{n\neq m}C_m^{*}C_nM_m^{*}(x)M_n(x)\langle E_m|E_n\rangle
\end{eqnarray}
can be obtained from the total wave function $|\Psi \rangle $ by ``summing
over'' all possible states of the environment. Here, $M_n(x)=\langle
x|M_n\rangle $ is the state of the macroscopic object in the position
representation. The second term on the r.h.s of the above equation is
responsible for the interference pattern of the macroscopic object. It is
easy to see that the interference fringes completely vanish when the states
of the environment are orthogonal to one another, i.e., when $\langle
E_m|E_n\rangle =\delta _{m.n}$. In this situation, an ideal ``which-way''
detection results from the ideal entanglement , in which one can distinguish
the states of environment very well. Our previous works on quantum
measurement theory [15,16] also show that an ideal entanglement may appear
in the macroscopic limit that the number $N$ of particles making up the
environment approaches infinity. It was found that the\ factorization
structure in the overlap integral $F_{m,n}=\langle E_m|E_n\rangle \equiv
\prod_{j=1}^N\ \langle E_m^{[j]}|E_n^{[j]}\rangle ,$where $F_{m,n}$ are the
overlapping of environment states, or the decoherence factor, plays the main
part in the quantum decoherence. Here, $|E_n^{[j]}\rangle $ are the single
states of those blocks constituting the environment. Since each factor $%
\langle E_m^{[j]}|E_n^{[j]}\rangle $ in $F_{m,n}$ has a norm less than
unity, the product of infinite such factors may approach zero. This
investigation was developed based on the Hepp-Coleman model[13].

Now we consider a macroscopic object with collective and internal variables,
say $x$ and $q$ . Applying the above discussion one easily sees that the
interaction between these two kinds of variables may lead to an ideal
quantum entanglement between the collective and internal states, when the
collective states are free of the back-action. This conception is initialed
by Onmes with the so-called ''internal environment''naming the system of
internal variables [9].But the question is whether the negligibility of the
back-action is the unique cause for the appearance of the above mentioned
''measurement-like process''. If not, what are the other causes beyond it?
To resolve this problem, we invoke the BO approximation to adiabatically
separate the collective and internal variables. Assume that the total
Hamiltonian is $H=\frac{p^2}{2M}+h(q,x)$, where the Hamiltonian $h(q,x)$
describes the motion of the internal variables $q$ coupling to the
collective variable $x$. For a fixed value of the slow variable $x$, the
eigen-state $|n[x]\rangle $ and the corresponding eigen-values $V_n[x]$ are
determined by the eigen- equation
\begin{equation}
h(q,x)|n[x]\rangle =V_n(x)|n[x]\rangle .
\end{equation}
Regarding $x$ and $q$ as the slow and fast variables respectively in the BO
adiabatic approach, we approximately obtain the complete set \{$\langle
x|n,\alpha \rangle \equiv \phi _{n,\alpha }(x)|n[x]\rangle $\} of
eigenstates of the total system, where $\phi _{n,\alpha }(x)$ come from the
eigen-equation
\begin{equation}
H_n\phi _{n,\alpha }(x)=E_{n,\alpha }\phi _{n,\alpha }(x)
\end{equation}
and
\begin{equation}
H_n=p^2/M+V_n[x]
\end{equation}
is the effective Hamiltonian associated with the internal state $%
|n[x]\rangle $. Here, we do not consider the induced gauge potential
connected with Berry phase factor through the quantum adiabatic method
[17,18]. Then, we can see how the ``measurement-like process'' naturally
appears as a result of the adiabatic dynamic evolution.

In fact, under the BO approximation, we can expand the factorized initial
state $|\Psi (0)\rangle =$ $|x\rangle \otimes |\phi \rangle $ in terms of
the adiabatic basis \{$|n,\alpha \rangle $\} and then we obtain the total
wave function[12]
\begin{equation}
|\Psi (t)\rangle ==\sum_n\langle n[x]|\phi \rangle \int dx^{\prime
}K(x^{\prime },x,t)|x^{\prime }\rangle \otimes |n[x^{\prime }]\rangle
\end{equation}
where we have used the completeness relations for the full eigen-functions
expressed in $x-representation$ and $K(x^{\prime },x,t)$ $=\langle x^{\prime
}|e^{-iH_nt}|x\rangle $ . Generally, the propagator $K(x^{\prime },x,t)$ is
not diagonal for $|x\rangle $ is not an eigen-state of $H_n$ and then $|\Psi
(t)\rangle $ can defines an ideal entanglement state. However, for the large
mass $M$ , we can prove that, to the first order approximation , $%
K(x^{\prime },x,t)$ takes a diagonal form proportional to a $\delta
-function.$ Actually, in the large limit, the kinetic term $p^2/2M$ can be
regarded as a perturbation in comparison with the effective potential $%
V_n(x).$ Using Dyson expansion to the first order of $\frac 1M$, we have
\[
e^{-iH_nt}=e^{-iV_nt}\left( 1-i\int_0^te^{iV_nt^{\prime }}\frac{p^2}{2M}%
e^{-iV_nt^{\prime }}dt^{\prime }+\cdots \right)
\]
\begin{equation}
=e^{-iV_nt}\left( 1-i\frac{p^2t^2}{2M}+i\frac{t^2}{4M}(p\partial
_xV_n+[\partial _xV_n]p)-\frac{it^3\partial _xV_n^2}{6M}+.......\right)
\end{equation}
Since
\[
\int \langle x^{\prime }|P^n|x\rangle f(x^{\prime })dx=0
\]
for n=1,2,..., we conclude that
\begin{equation}
K(x^{\prime },x,t)=e^{-iV_n[x]t}[\delta (x-x^{\prime })+\frac
i{2M}\int_0^td\tau e^{-iV_n[x^{\prime }]\tau }\frac{\partial ^2}{\partial
x^{\prime 2}}\delta (x-x^{\prime })e^{iV_n(x)\tau }].
\end{equation}
Then, we show that if is approximately diagonalized: $K(x^{\prime },x,t)=$ $%
e^{-iH_n(x)t}\delta (x-x^{\prime })$ , the adiabatic wave-function can lead
to an ideal entanglement

\begin{equation}
|\Psi (t)\rangle =\sum_n\langle n[x]|\phi \rangle e^{-iH_n(x)t}|x\rangle
\otimes |n[x]\rangle
\end{equation}
We call this entanglement adiabatic entanglement.

In conclusion , up to the first order approximation of $\frac 1M$, we have
the quantum entanglement in adiabatic evolution. The Born- Oppenheimer
adiabatic approximation has provided us with a novel mechanism to produce
the quantum entanglement between the macroscopic object and its internal
variables.

\section{Localization Induced by Factorized Internal Motion}

We notice the above simple result has the following physical explanation:
the evolution state of a heavy particle for very large $M$, which is almost
steady, is approximately an eigenstate of the position operator if it is
initially in a state with a fixed position. Then, it follows from eq.(13)
that, in the large-mass limit, the wave function $|\Psi (t)\rangle $ can be
factorized approximately: $|\Psi (t)\rangle =|x\rangle \otimes S(x,t)|\phi
\rangle $ where the entangling $S-matrices$%
\begin{equation}
S(x,t)=\sum_{n,}e^{-iV_nt}|n[x]\rangle \langle n[x]|
\end{equation}
are defined in terms of the adiabatic projection $|n[x]\rangle \langle n[x]|$%
.

According to our previous argument about the factorized structure of $%
S-matrix$ in the dynamic theory of quantum measurement [11,12 ], if the
internal degree of freedom has many components, e.g.,if $q=(q_1,q_2,...q_N)$
, then in their normal non-interaction modes , $S(x;t)$ can be factorized
as:
\begin{equation}
S(x;t)=\prod_{j=1}^NS_{_j}(x;t)
\end{equation}
with
\begin{equation}
S_{_j}(x;t)=e^{-ih_j(q_j,x)t}
\end{equation}
with $h(q,x)=\sum_jh_j(q_j,x)$ and $h_j(q_j,x)$ are the single particle
Hamiltonians of the parts of the macroscopic object. Of course., in the
derivation of the above factorized structure for the $S-matrix$ , we have
made some simplifications. Roughly speaking, we have assumed that the
adiabtaic effective potential takes the form of direct sum $%
V_n=\sum_jV_{nj}(q_j),$and the eigenstate the form of direct product

\begin{equation}
\qquad |n[x]\rangle =\prod_{j=1}^N\otimes |n_j[x]\rangle
\end{equation}
neglecting the higher order terms $\approx $ $O(\frac 1M).$

For the initial state $|\phi \rangle =\prod_{j=1}^N\otimes $ $|\phi
_j\rangle $ factorized with respect to internal components, the reduced
density matrix
\begin{equation}
\rho (x,x^{\prime },t)=\varphi (x)\varphi ^{*}(x^{\prime })F_N(x^{\prime
},x,t):
\end{equation}
can be re-written in terms of the so called decoherence factor
\begin{equation}
F_N(x^{\prime },x,t)=\prod_{j=1}^NF^{[j]}(x^{\prime },x,t)\equiv
\prod_{j=1}^N\langle \phi _j|S_{q_j}^{\dagger }(x^{\prime
};t)S_{q_j}(x;t)|\phi _j\rangle .
\end{equation}
This factor is expressed as an $N$-multiple product of the single decohering
factors
\begin{equation}
F^j(x,x^{\prime })=\langle \phi _j|S_{q_j}^{\dagger }(x^{\prime
};t)S_{q_j}(x;t)|\phi _j\rangle
\end{equation}
with norms less than unity. Thus in the macroscopic limit $N\rightarrow
\infty $ , it is possible that $F_N(x^{\prime },x,t)$ $\rightarrow 0,$ for $%
x^{\prime }\neq x$. In fact, this factor reflects almost all the dynamic
features of the influence of the fast part on the slow part. Physically, an
infinite $N$ means that the object is macroscopic since it is made of
infinite number of particles in that case. On the other hand, the happening
of decoherence at infinite $N$ manifests a transition of the object from the
quantum realm to the classical realm. Here,as expected,the physical picture
is consistent.

As to the localization problem raised by Einstein and Born [1], we , based
on the above argument, comment that one can formally write down the wave
function of a macroscopic object as an narrow pure state wave packet, but it
is not the whole of a real story. Actually, the statement that an object is
macroscopic should physically imply that it contains many particles. So a
physically correct description of its state must concern its internal
motions coupling to the collective coordinates (e.g., its center-of-mass) .
Usually, one observe this collective coordinate to determine whether two
spatially-localized wave packets can interfere with each other. If there
does not exist such interference, one may say that, the superposition of two
narrow wave packets for the macro-coordinate is no longer a possible pure
state of the macroscopic object. Indeed, because the ``which-way''
information of the macro-coordinate is recorded by the internal motions of
particles making up the macroscopic object, the induced decoherence must
destruct the coherence in the original superposition so that the state of
the macroscopic object is no longer pure.

The present argument also provides a possible solution for the Schroedinger
cat paradox. If we consider the Schroedinger cat as a macroscopic object
consisting of many internal particles, then we can never observe anything
corresponding to the interference between the dead and the living cats .
This is because the macroscopically- dead and the macroscopically- living
states, $|D\rangle $ and $|L\rangle ,$ of the cat are correlated to the
corresponding internal states, $|d_j\rangle $ and $|l_j\rangle $. The cat
state
\begin{equation}
|Cat\rangle =|L\rangle \otimes \prod_{j=1}^N|l_j\rangle +|D\rangle \otimes
\prod_{j=1}^N|d_j\rangle
\end{equation}
follows from the argument this section when $|D\rangle $ and $|L\rangle $
are regarded as the collective states while $\prod_{j=1}^N|l_j\rangle $ and $%
\prod_{j=1}^N|d_j\rangle $ describes the corresponding internal motion. It
leads to a reduced density matrix with the off-diagonal elements
proportional to $\prod_{j=1}^N\langle d_j|l_j\rangle .$ once there is only
one pair of inner states are orthogonal, the off-diagonal elements vanish
and decoherence happen. Even though there does not exists any pair of inner
states orthogonal with each other, for the norm of $\langle d_j|l_j\rangle $
less than unity, it is also possible that $\prod_{j=1}^N\langle
d_j|l_j\rangle \rightarrow 0$ in the macroscopic limit $N\rightarrow \infty .
$ In this sense, we conclude that the Schroedinger cat paradox is not a
paradox at all in practice.. Rather, it essentially arises from overlooking
the internal motions of a macroscopic cat or the multi-particle scattering
off it.

Now, we have shown that the localization phenomena of a macroscopic object
can boil down to an entanglement between its collective position (or C.M)
and internal variables in the adiabatic evolution with the above mentioned
factorization structure. Closely related to the Schroedinger cat phenomenon,
this entanglement results from their adiabatic separation of collective and
internal variables.With the point view from the above theoretical analysis,
as for the C$_{60}$ molecule interference experiment. To our surprise, it
turns out that an elegant interference pattern appears in the experiment.
But there is no contradiction here. Truly, at high temperature C$_{60}$
would emit two or three infrared photons during its passage through the
apparatus. But as the wavelength of the radiating photons from the internal
motion of C$_{60}$ is much greater than the distance between the neighboring
slits, the photons carry no information about the route the molecule takes.
In this sense, C$_{60}$ can not well be considered as a macroscopic object
since its internal variable is almost frizzed.. Therefore the interference
pattern is not affected. Similarly, though there exists interaction with the
external air particles, the scattering rates on the macroscopic object are
far too small to induce quantum decoherence. This explains the persistence
of interference pattern in the experiment [10]. However, we can imagine that
in such experiments, the internal motion (such as radiation of photons of
various frequencies) produces an effective coupling with the collective
motion of the C.M. Then, the configurations of internal motion can record
the ''which-way'' information even through a single thermal photon so that
the interference contrast should thus be completely destroyed. Moreover, the
parameters (such as the internal temperature of the fullerenes, the
temperature of the environment, the intensity and frequency of external
laser radiation ) can be controlled continuously so that quantitative
natures such as those described in this paper could be tested. The study in
the present paper is not directly applicable to such ''which-way''
experiments because it is based on the assumption that the macroscopic
object is composed of two-level subsystems and does not concern the concrete
structure of C$_{60}$ fullerenes. Nevertheless, for the quantitative
investigation of the dynamic details of decoherence process in such
experiments, it can serve as a starting point.

\section{Simple Model for Macroscopic Localization}

To make a deeper elucidation of the above general arguments about the
localization of a macroscopic object of mass $M$, we model the macroscopic
object as consisting of $N$ two level particles, which are fixed at certain
positions to form a whole without internal spatial motion. The collective
position $x$ is taken to be its mass-center or any reference position in it
while the internal variables are taken to be the quasi-spins associated with
two level particles. Generally, if we assume that the back-action of the
internal variables on the collective position is relatively small, the model
Hamiltonian can be written as
\[
H=\frac{P^2}{2M}+h(x):
\]

\begin{equation}
h(x)=\sum_{j=1}^N(f_j(x)|e_j\rangle \langle g_j|+f_j^{*}(x)|g_j\rangle
\langle e_j|)+\sum_{j=1}^N\omega _j(|e_j\rangle \langle e_j|-|g_j\rangle
\langle g_j|)
\end{equation}
where $|g_j\rangle $ and $|e_j\rangle $ are the ground and the excited
states of the $j$ 'th particle and $f_j(x)$ denotes the position-dependent
couplings of the collective variable to the internal variables. Let $l_j$ be
the relative distance between the $j$ 'th particle and the reference
position $x.$ Further we assume $f_j(x)=f(x+l_j).$ Physically,we may think
that these couplings are induced by an inhomogeneous external field , e.g.,
they may be the electric dipole couplings of two-level atoms in an
inhomogeneous electric field.

We remark that the above model enjoys some universality under certain
conditions, compared with various environment models inducing both
dissipation and decoherence of quantum processes. In fact, Caldeira and
Leggett [19] have pointed out that any environment weakly coupling to a
system may be approximated by a bath of oscillators under the condition that
``each environmental degree of freedom is only weakly perturbed by its
interaction with the system''$.$We observe that any linear coupling only
involves transitions between the lowest two levels (ground state and the
first excitation state) of each harmonic oscillator in the perturbation
approach though it has many energy levels. Therefore in such a case we can
also describe the environment as a combination of many two level subsystems
without losing generality [20].To some extent, these arguments justify our
choosing the two level subsystems to model the internal motion of the
macroscopic object.We will soon see its advantage:the character of
localization can be manifested naturally and clearly.

Now let us calculate the $S_{_j}(x;t)$ for this concrete model. The
single-particle Hamiltonian $h_j(x)=\omega _j(|e_j\rangle \langle
e_j|-|g_j\rangle \langle g_j|)+(f_j(x)|e_j\rangle \langle g_j|+h.c)$ has the
$x$-dependent eigenvalues
\begin{equation}
V_{jc}=n\Omega _j(x)\equiv \pm \sqrt{|f_j(x)|^2+\omega _j^2}\qquad (n=\pm )
\end{equation}
and the corresponding eigen-vectors$|n_j[x]\rangle $ are
\begin{equation}
|+_j[x]\rangle =\cos \frac{\theta _j}2|e_j\rangle +\sin \frac{\theta _j}%
2|g_j\rangle ,
\end{equation}
\begin{equation}
|-_j[x]\rangle =\sin \frac{\theta _j}2|e_j\rangle -\cos \frac{\theta _j}%
2|g_j\rangle ,
\end{equation}
where $\tan \theta _j=\frac{f_j(x)}{\omega _j}.$ Explicitly, the
corresponding single-particle $S-matrix$

\begin{equation}
S_{_j}(x;t)=\left(
\begin{array}{cc}
\cos (\Omega _jt)-i\sin (\Omega _jt)\cos \theta _j, & i\sin (\Omega _jt)\sin
\theta _j \\
i\sin (\Omega _jt)\sin \theta _j, & \cos (\Omega _jt)+i\sin (\Omega _jt)\cos
\theta _j
\end{array}
\right)
\end{equation}
Here in the derivation we have used the formula
\begin{equation}
\exp [i\overrightarrow{\sigma }\cdot \overrightarrow{A}]=\cos A+i%
\overrightarrow{\sigma }\cdot \overrightarrow{n_A}\sin A
\end{equation}
for a given vector $\overrightarrow{A}$ of norm $A$ along the direction $%
\overrightarrow{n_A}.$Having obtained the above analytic results about $%
S-matrix,$ we can further calculate the single-particle decoherence factors $%
F^{[j]}(x^{\prime },x,t)\equiv \langle g_j|S_{_j}^{\dagger }(x^{\prime
};t)S_{_j}(x;t)|g_j\rangle $ for a given initial state $|\phi \rangle
=\prod_{j=1}^N\otimes $ $|g_j\rangle $. For simplicity we use the notation $%
f(x^{\prime })=f^{\prime }$ .We have
\[
F^{[j]}(x^{\prime },x,t)=\{\sin (\Omega _j^{\prime }t)\sin \theta _j^{\prime
}\sin (\Omega _jt)\sin \theta _j+
\]
\begin{equation}
\cos (\Omega _j^{\prime }t)\cos (\Omega _jt)+\sin (\Omega _j^{\prime }t)\cos
\theta _j^{\prime }\sin (\Omega _jt)\cos \theta _j^{\prime }\cos \theta _j
\end{equation}
\[
+i\{\cos (\Omega _j^{\prime }t)\sin (\Omega _jt)\cos \theta _j-\sin (\Omega
_j^{\prime }t)\cos \theta _j^{\prime }\cos (\Omega _jt)\}\}
\]
In the weakly coupling limit, $g_j\ll \omega _j$ and the coupling $f_j\simeq
g_jx$, thus we have $\sin \theta _j\simeq \theta _j\simeq \frac{f_j}{\omega
_j},\cos \theta _j\simeq 1-\frac 12\theta _j^2$ and $\Omega _j\simeq \omega
_j.$Then, the decohering factors can be simplified as
\begin{equation}
F^{[j]}(x^{\prime },x,t)\simeq 1-(x-x^{\prime })^2\frac{|g_j|^2}{2\omega _j^2%
}\sin ^2(\omega _jt) \\[-4mm]
+\frac{i|g_j|^2}{4\omega _j^2}(x^2-x^{\prime 2})\sin (2\omega _jt)
\end{equation}
Consequently, the temporal behavior of the decoherence is determined by
\begin{equation}
F(x^{\prime },x,t)=\left| F(x^{\prime },x,t)\right| \exp (\frac{i|g_j|^2}{%
4\omega _j^2}(x^2-x^{\prime 2})\sin (2\omega _jt))
\end{equation}
where
\begin{equation}
\left| F(x^{\prime },x,t)\right| =\exp (-(x-x^{\prime })^2\frac{|g_j|^2}{%
2\omega _j^2}\sin ^2(\omega _jt))
\end{equation}

In the case of continuous spectrum, the sum
\begin{equation}
R(t)=\sum\limits_{j=1}^N\frac{g_j^2}{2\omega _j^2}\sin ^2\left( \omega
_jt\right)
\end{equation}
can be re- expressed in terms of a spectrum distribution $\rho (\omega _k)$
as
\[
R(t)=\int_0^\infty \frac{\rho (\omega _k)g_k^2}{2\omega _k^2}\sin ^2\omega
_kd\omega _k.
\]

From some concrete spectrum distributions, interesting circumstances may
arise. For instance, when $\rho (\omega _k)=\frac 4\pi \gamma /g_k^2$ the
integral converges to a negative number proportional to time t , precisely, $%
R(t)=\gamma t$ .Therefore, our analysis recovers the result
\begin{equation}
\rho (x,x^{\prime },t)=\varphi (x)\varphi ^{*}(x^{\prime })e^{-\gamma
t(x-x^{\prime })^2}\exp [i\pi (x^2-x^{\prime 2}{})s(t)]
\end{equation}
for the reduced density matrix of the macroscopic object , which was
obtained by Joos and Zeh [3] through the multi particle external scattering
mechanism and by Zurek separately through Markov master equation. Here,
\begin{equation}
s(t)=\sum_{j=1}^N\frac{\sin (2\omega _jt)}{4\pi \omega _j^2}
\end{equation}
is a time-dependent periodic function. This shows that the norm of the
decoherence factor is exponentially decaying and as $t\rightarrow \infty ,$
the off-diagonal elements of the density matrix vanish simultaneously!

We will show that for quite general distribution $\rho (\omega )$ the
off-diagonal elements of the reduced density matrix decline rather sharply
with time $t$ if the particle number $N$ is large. Assume that all $%
g_j^{\prime }s$ are equal:$g_j=g.$ If the frequencies lie within an interval
$[\omega _1,\omega _2]$ and the distribution is homogeneous, we have $\rho
(\omega )=N/(\omega _2-\omega _1).$ Then $\overline{}$%
\begin{eqnarray}
R(t) &=&\int_{\omega _1}^{\omega _2}\frac{g^2}{2\omega ^2}\sin ^2\omega t\
\rho (\omega )d\omega   \nonumber \\
&=&\frac N{(\omega _2-\omega _1)}\frac{g^2}2\int_{\omega _1}^{\omega
_2}\frac 1{\omega ^2}\sin ^2\omega t\ d\omega   \nonumber \\
&\geqslant &\frac N{(\omega _2-\omega _1)}\frac{g^2}{2\omega _2^2}%
\int_{\omega _1}^{\omega _2}\sin ^2\omega t\ d\omega   \nonumber \\
&=&\frac N4\frac{g^2}{\omega _2^2}\left( 1-\cos (\omega _2+\omega _1)t\frac{%
\sin (\omega _2-\omega _1)t}{(\omega _2-\omega _1)t}\right)
\end{eqnarray}
For a general $\rho (\omega )$ in the interval $[\omega _1,\omega _2],$ we
have
\[
\int_{\omega _1}^{\omega _2}\rho (\omega )d\omega =N.
\]
Then there exists some $\overline{\omega }$ in $[\omega _1,\omega _2]$ such
that
\[
\rho (\overline{\omega })=\frac N{\omega _2-\omega _1}
\]
If the frequency spectrum of the system is such that there exist $\omega _3$
and $\omega _4$ in the interval $[\omega _1,\omega _2]$ satisfying
\begin{equation}
\rho (\omega )\geqslant \frac N{\omega _2-\omega _1}\text{ for }\omega
_3\leqslant \omega \leqslant \omega _4.
\end{equation}

From the derivation of (44) it then follows that
\begin{equation}
R(t)\geqslant \frac N4\frac{g^2}{\omega _4^2}\frac{\omega _4-\omega _3}{%
\omega _2-\omega _1}\left( 1-\cos (\omega _4+\omega _3)t\frac{\sin (\omega
_4-\omega _3)t}{(\omega _4-\omega _3)t}\right)
\end{equation}
After a moment's thought, one can easily convince oneself that the condition
(36) is rather easy to satisfy. From the inequality (37) we observe that
although in the weakly coupling limit, we should have $\frac{g^2}{\omega _4^2%
}\ll 1,$ $R(t)$ can increase sharply with time $t$ if the particle number is
large enough. This just means that the off-diagonal elements of the reduced
density matrix will decline sharply with time $t.$ In conclusion, despite
the complexity of $\rho (x,x^{\prime },t)$ due to the presence of the
oscillating factor $s(t)$, in many cases it can well describe the
decoherence of macroscopic object thanks to its simple decaying norm.

Let us now turn to consider an example similar to that studied by Joos and
Zeh .We take a coherent superposition of two Gaussian wave packets of width $%
d$
\begin{equation}
\varphi (x)=\frac 1{\sqrt[4]{8\pi d^2}}\left\{ \exp \left( -\frac{(x-a)^2}{%
4d^2}\right) +\exp \left( -\frac{(x+a)^2}{4d^2}\right) \right\}
\end{equation}
The norm of the corresponding reduced density matrix
\begin{equation}
|\rho (x,x^{\prime },t)|=\sum_{k,l=0}^1P_{kl}(x,x^{\prime },t)
\end{equation}
contains 4 peaks:
\[
P_{11}(x,x^{\prime },t)=\frac 1{\sqrt{8\pi d^2}}e^{-\gamma t(x-x^{\prime
})^2}\exp [-\frac{(x-a)^2}{4d^2}-\frac{(x^{\prime }-a)^2}{4d^2}]
\]
\[
P_{10}(x,x^{\prime },t)=\frac 1{\sqrt{8\pi d^2}}e^{-\gamma t(x-x^{\prime
})^2}\exp [-\frac{(x-a)^2}{4d^2}-\frac{(x^{\prime }+a)^2}{4d^2}]
\]
\begin{equation}
P_{01}(x,x^{\prime },t)=\frac 1{\sqrt{8\pi d^2}}e^{-\gamma t(x-x^{\prime
})^2}\exp [-\frac{(x+a)^2}{4d^2}-\frac{(x^{\prime }-a)^2}{4d^2}]
\end{equation}
\[
P_{00}(x,x^{\prime },t)=\frac 1{\sqrt{8\pi d^2}}e^{-\gamma t(x-x^{\prime
})^2}\exp [-\frac{(x+a)^2}{4d^2}-\frac{(x^{\prime }+a)^2}{4d^2}]
\]
centering respectively around the points $(a,a),(a,-a),(-a,a)$ and $(-a,-a)$
in $x-x^{\prime }$-plane. The heights are respectively $1/\sqrt{8\pi d^2}%
,e^{-4\gamma ta^2}/\sqrt{8\pi d^2},e^{-4\gamma ta^2}/\sqrt{8\pi d^2}$and $1/%
\sqrt{8\pi d^2})$. Obviously, two peaks with centers at $(a,-a)$ and $(a,-a)$
decay with time while the other two keep their heights constant. Fig.1.
\begin{figure}[h]
\begin{center}
\includegraphics[width=7in,height=6in]{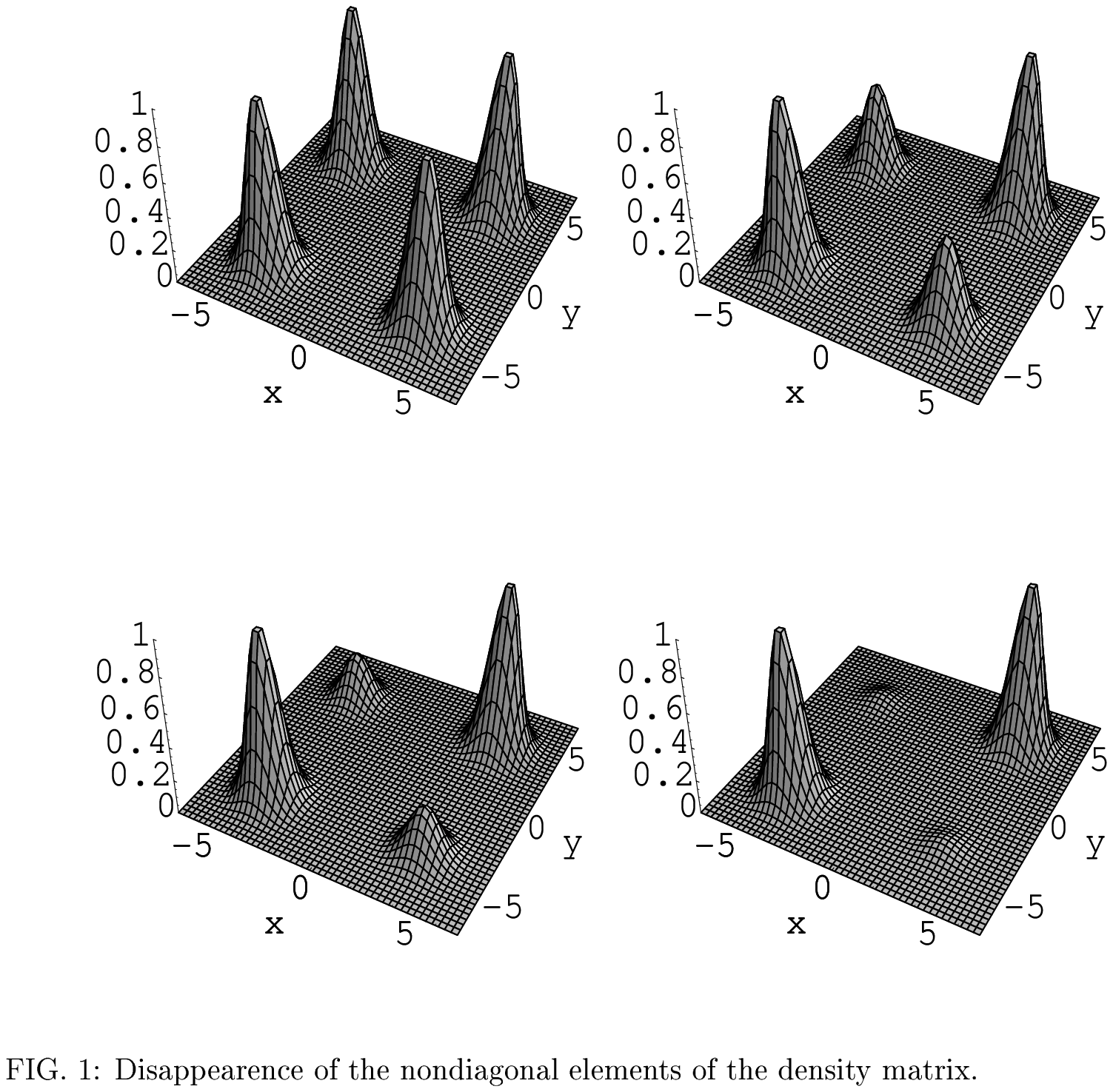}
\end{center}
\end{figure}
shows this time-dependent configuration at t=0,and a finite t. As $%
t\rightarrow \infty $, two off-diagonal terms $P_{10}$ and $P_{01}$decay to
zero so that the interference of the two Gaussian wave packets are destroyed
. In this sense, we say that the pure state $\rho (x,x^{\prime },t=0)=\int
dx\varphi (x)\varphi *(x^{\prime })|x\rangle \langle x^{\prime }|$ becomes a
mixed state
\begin{equation}
\rho (t)=\int dx\varphi (x)\varphi ^{*}(x)|x\rangle \langle x|
\end{equation}
in $x-$representation.\qquad

Interference of two plane waves of wave vector $k_1,k_2$ provides us another
simplest example. Without decoherence induced by its internal motions or the
external scattering , their coherent superposition $\varphi (x)=\sqrt{\frac
1{4\pi }}[e^{ik_1x}+e^{ik_2x}]$ yields a spatial interference described by
the reduced density matrix
\[
\rho _0(x,x^{\prime },t)=\frac 1{4\pi }\{e^{ik_1(x-x^{\prime
})}+e^{ik_2(x-x^{\prime })}+
\]
\begin{equation}
\exp [i(\frac{k_1^2t-k_2^2t}{2m}+k_2x-k_1x^{\prime })]+\exp [i(\frac{%
k_2^2t-k_1^2t}{2m}+k_1x-k_2x^{\prime }]\}
\end{equation}
Under the influence of internal motions , it becomes
\[
\rho (x,x^{\prime },t)\approx \rho _0(x,x^{\prime },t)e^{-\gamma
t(x-x^{\prime })^2}
\]
for large mass. We see that the difference created by decoherence is only
reflected in the off-diagonal elements,and the pure decoherence (without
dissipation) does not destroy the interference pattern described by the
diagonal term $\rho (x,x,t)=\rho _0(x,x,t).$This simple illustration tells
us that the present quantum decoherence mechanism may not have to do with
the interference pattern of the first order coherence, but it does destroy
the higher order quantum coherence: $\rho (x,x^{\prime },t)\rightarrow 0$ as
$t\rightarrow \infty .$ In fact, due to the induced loss of energy, quantum
dissipation is responsible for the disappearance of the interference pattern
of the first order coherence. The influences of internal motions or external
scattering on the decoherence of a macroscopic object may be very
complicated. Intuitively, these dynamic effects should depend on the details
of interaction between the collective variables and the internal and
external degrees of freedom. Practically,we can classify these influences
into two categories, namely, quantum dissipation and quantum decoherence,
and then study them separately by different models.

\quad

\section{Concluding Remarks}

It is noticed that, so long as the `` which-way'' information of the
collective motion of a macroscopic object already stored in the internal
motion could be read out, the phenomenon caused by interference would be
destroyed without any data being read out in practice [5,13-16]. In this
sense the internal degrees of freedom interacting with the macroscopic
object behaves as a detector to realize a ``measurement-like '' process.
Thus, the internal motion configuration is imagined as an objective detector
detecting the collective states . Provide the internal motion configuration
couples with the collective motion and produce an ideal entanglement, the
collective motion must lose its coherence. It is worthy to point out that
this simple entanglement conserves the energy of the collective motion while
destroying the quantum coherence.

In the case with no energy conservation, the quantum dissipation can also
induce the localization of the macroscopic object. Based on the studies of
quantum dissipation stimulated by Caldeira and Leggett [18], Yu and one
(C.P.S) of the authors found a novel mechanism which sheds new light on the
localization problem of macroscopic object [21]. They studied the quantum
dynamics of a simplest dissipative system: one particle moving in a constant
external field and interacting with a bath of harmonic oscillators with
Ohmic spectral density. It was found that the wave function of the total
system can be factorized as a product of the system part and the bath part.
When one ignores the effect of Brownian motion or the quantum fluctuation in
the system caused by the bath, the product wave function becomes a direct
product and the dissipative evolution of the system is governed by
Caldirora-Kani (CK) Hamiltonian . Using this effective Hamiltonian, they
discovered the following interesting result: the dissipation suppresses the
wave packet spreading and cause the localization of the wave packet.
Actually,it was shown that the breadth of the wave packet changes with time $%
t$ in the following way: $w(t)=a\sqrt{1+\frac{t_\eta ^2}{4M^2a^4}.}$Here $a$
is the initial breadth of the wave packet and $t_\eta ={\frac{M(1-e^{-\eta
t/M)}}\eta },$where $\eta $ is the damping rate. Comparing this formula with
the equation (1), we find that the effect of the influence of the bath is
the replacement of $t$ by $t_\eta $ in (1). We have $t_\eta \rightarrow t$
when $\eta /M\rightarrow 0$. So one can regard $t_\eta $ as a deformation of
time $t$ caused by dissipation. Notice that $t_\eta $ approaches the limit $%
\frac M\eta $ as $t\rightarrow \infty $. This means localization of the wave
packet in the presence of dissipation. Indeed, we have the limit breadth:$%
a_{limit}=a\sqrt{1+(1/2\eta a^2)^2}.$This suppression of the wave packet
spreading by dissipation possibly provides a useful mechanism for the
localization of quantum particle. It is a little bit surprising that the
limit width of the damped particle wave packet as $t\rightarrow \infty $ is
exactly the same as the ``uncertainty product'' of the damped particle,
established by Schuch et al. through nonlinear Schroedinger equation [22] .

Summarily, the environment induced dissipation as well as decoherence can
provide an important mechanism for the localization of a macroscopic object.
Mentioning macroscopicness implies the requirement that the macroscopic
object must contain a large number of internal blocks. Then the macroscopic
object, coupled to the internal variables, should be described by collective
variables subject to an interaction similar to that concerning the external
scattering in WJZ mechanism and the quantum dissipation of a particle in a
bath.

\section*{Acknowlegement}

\noindent This work is supported by the NFS of China and in part by the
Erwin Schroedinger Institute, Vienna. One of the authors (CPS) wishs to
express his sincere thanks to for many useful discussions with J.W.Pan, Y.Wu
and L.M.Kuang.

\end{document}